# Survey of charging scheduling, fleet management, and location planning of charging stations for electrified demand-responsive transport systems: methodologies and recent developments

**Tai-Yu Ma* and Yumeng Fang**

Luxembourg Institute of Socio-Economic Research (LISER), Department of Urban Development and Mobility, Esch-sur-Alzette, Luxembourg

*Corresponding author

**Abstract**

The accelerated electrification of transport systems with EVs has brought new challenges for charging scheduling, fleet management, and charging infrastructure location and configuration planning. In this review, we have provided a systematic review of the recent development in strategic, tactical, and operational decisions for demand responsive transport system planning using electric vehicles (EV-DRT). We have summarized recent developments in mathematical modeling approaches and identified future research directions. A list of existing open-access datasets, numerical test instances, and software are provided for future research in EV-DRT and related problems.

## 1. Introduction

Since 1970, demand-responsive transport (DRT) has received broad interest as an efficient alternative to improve the accessibility and coverage of fixed-route public transport in low-density areas (Volinski, 2019). DRT covers a spectrum of services that can be operated as door-to-door services, feeder services connecting to transit stations, or flexible bus services using point/route-deviation strategies (Chow et al., 2020). Users book their ride requests in advance via dedicated apps and platforms, and operators can design their services to adapt to user demand. An increasing number of public transport agencies have launched DRT pilots to meet users' needs in low-demand areas. It has been shown that integrating DRT as a feeder service could increase the ridership of transit and reduce congestion and $CO_2$ emissions (Hazan et al., 2020).

In the context of the current climate crisis, the transport sector faces an unprecedented challenge in terms of the transition to clean energy. The transport sector contributed to 27% of total EU-27 emissions in 2017 (EEA, 2021), and in order to meet the EU's climate-neutral target the transport sector needs to reduce its emissions by about two-thirds by 2050. In the face of this challenge, transport network companies (TNCs) need to carefully analyze charging infrastructure and fleet requirements and develop strategies to minimize operation costs. First, the battery capacity of electric vehicles (EVs) is limited (to about 100–300 miles[1]), and within-day charging operations are necessary. Second, charging times are long (i.e., generally 3–12 hours, although only 30 minutes are required for an 80% charge when using a fast charger[1]) and public DC fast chargers are still rare in many cities due to their high investment cost. Recent studies on the impact of the electrification of ride-hailing services in the USA show that TNCs need to recharge their vehicles several times a day and mainly rely on DC fast chargers to minimize charging times (Jenn, 2019). However, using DC fast chargers increases charging costs by up to 25% (Pavlenko et al., 2019). From the perspective of TNCs, the transition to EV-DRT requires developing adequate planning and management strategies, which can be grouped into three decision levels: strategic (charging infrastructure), tactical (fleet size), and operational (charging scheduling and routing) decisions. At a higher system level, the interactions of charging operations with the power grid also need to be considered in order to enhance the stability of the power grid. Recent literature reviews have summarized some of the methodologies used to address these issues. For example, Shen et al. (2019) provide a literature review of state-of-the-art mathematical modeling approaches for EV charging scheduling and the charging-infrastructure planning of car-sharing systems. Rahman et al. (2016) review different charging systems and optimization models for the charging-infrastructure planning of plug-in hybrid EVs and non-hybrid EVs. For electric buses, Olsen (2020) provides an overview of different mathematical modeling approaches for the charging scheduling and location planning of electric bus systems. Deng et al. (2021) focus on the different technologies used for energy storage, power management, and charging scheduling of electric bus systems. While the literature is large, there is still no systematic overview covering the state-of-the-art methodologies at the different levels of decision-making for EV-DRT systems. This review aims to fill this gap, considering EV-DRT systems including ride-sharing, ride-hailing, flexible buses, and other forms of on-demand transportation systems using a fleet of EVs or e-buses.

[1] https://fueleconomy.gov/feg/evtech.shtml

The main contributions of this review are as follows:

- Provide an in-depth literature review of recent developments in the mathematical modeling of dynamic EV-DRT system operational policy design and optimization;
- Review state-of-the-art methodologies and solution algorithms for the strategic and tactical planning of EV-DRT systems;
- Survey the datasets, test instances, and software used for EV-DRT system planning and research;
- Identify current research gaps and future research directions.

The remainder of this paper is organized as follows. In Section 2 we present the characteristics of EV-DRT systems, including charging operations, and discuss the different decision-making problems at the strategic, tactical, and operational levels. Section 3 provides a literature review and discusses recent developments related to these decision problems, focusing on the modeling approaches and solution algorithms employed. Section 4 surveys publicly available datasets, test instances, and software. Finally, we identify research gaps, discuss future research directions, and offer some concluding remarks.

## 2. Characteristics of EV-DRT systems and charging operations

### 2.1. System characteristics of EV-DRT systems

EV-DRT systems are demand-driven, reservation-based passenger transport services implemented in low-density rural/dense urban areas using a fleet of EVs. They provide either door-to-door services or serve as feeder services to connect transit stations as a part of multimodal mobility solutions. Different from conventional DRT using internal combustion engine vehicles, EVs are constrained by their limited battery range and require charging at depots or public charging stations. Depending on the charging power and charging technologies, the charging times and installation costs of charging stations vary significantly. For example, a Volkswagen Golf with a 300 km range requires 45 minutes to charge to 80% using a 50 kW DC fast charger and 10 hours when using a 3.6 kW charger (Spöttle et al., 2018). The unitary charger purchase cost (not including installation costs) ranges from around 800 euros for AC mode 2 home chargers to 40,000–60,000 euros for a DC fast charger with 100–400 kW of power (Spöttle et al., 2018). Due to the high investment cost of high-powered DC fast chargers, there are very few public DC fast-charging points, which is the main obstacle for DRT system electrification. To manage this challenge, TNCs need to develop efficient charging management strategies and infrastructure so as to optimize their charging operations while meeting their service-quality commitments. Unplanned charging operations result in higher charging operation costs, higher vehicle idle times for recharging, and significant ridership losses. Figure 1 presents these decision problems and their dependence on the context, including the relevant energy market and energy network, local demand, and the existing transport network.

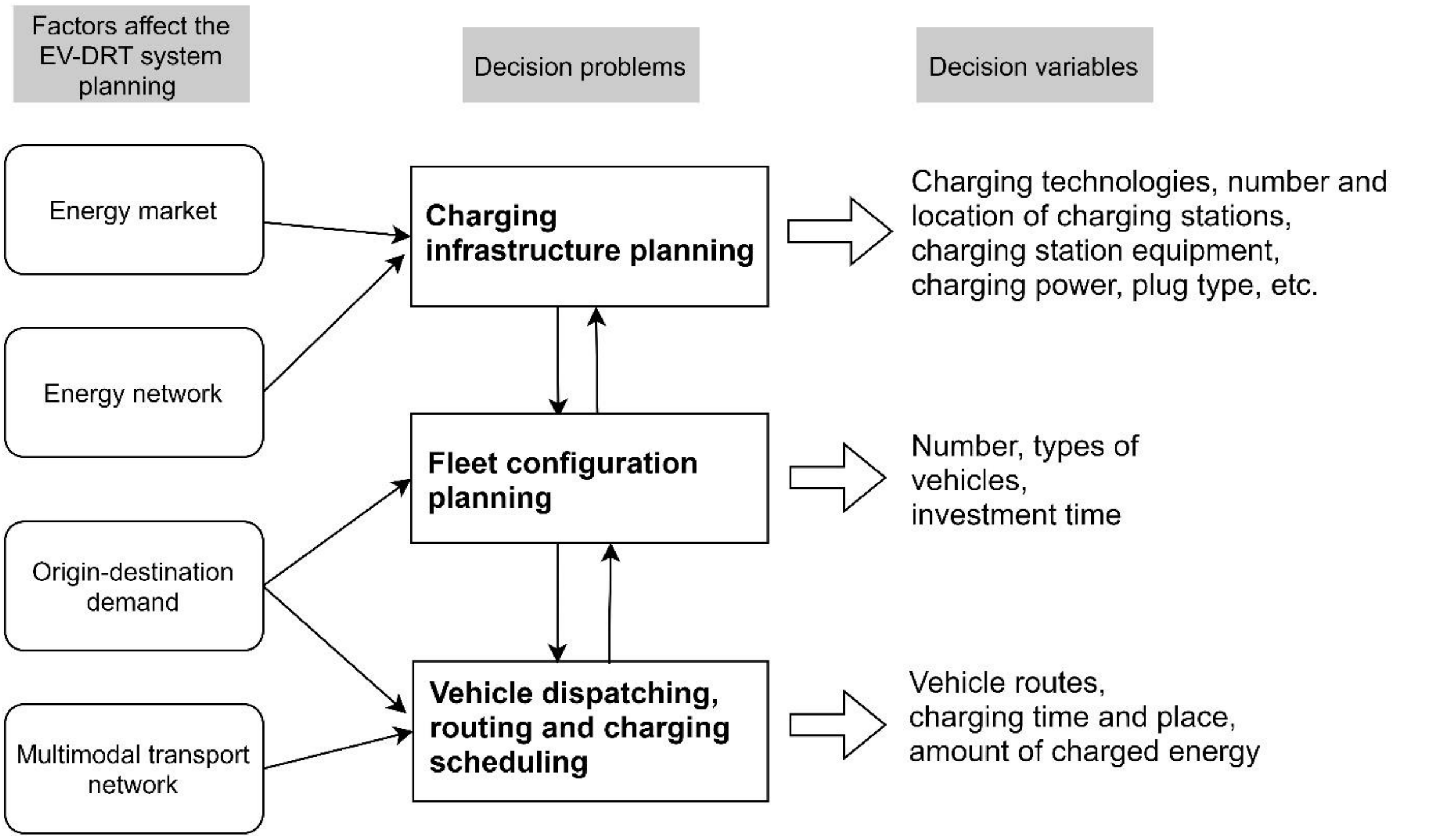

Figure 1. Strategic, tactical, and operational decisions for EV-DRT system planning and their dependence on the context.

The characteristics of the strategic, tactical, and operational decisions for EV-DRT system planning are summarized as follows.

- Operational decisions: This decision problem involves optimizing the fleet's daily charging scheduling to minimize charging costs while satisfying charging infrastructure constraints and customer demand over a short-term horizon (daily decisions). The problem is generally considered in an EV-based vehicle-routing problem (VRP) framework to handle additional constraints related to vehicle state of charge (SOC) using a simple needs-based charging policy (i.e., visiting charging stations when a vehicles' SOC is lower than a threshold (see the recent review of EV-VRPs by Kucukoglu et al., 2021). In the following, we review the existing literature related to modeling aspects of the charging process, energy consumption, charging policy, charging cost, and waiting time at charging stations.
  a. Battery energy consumption and charging function

EV battery energy discharge/consumption depends on numerous factors such as vehicle speed, load, road gradient, temperature, and acceleration/breaking, etc. (Fiori et al., 2016). From an operational optimization perspective, the energy consumption of EVs is generally assumed to be a linear function of traveled distance (Chen et al., 2016a; Ma, 2021, among many others). In terms of battery charging, the battery's SOC increases linearly from empty to a critical point with a constant charging rate, and then the charging rate decreases asymmetrically until the vehicle is fully charged. Zalesak and Samaranayake (2021) propose an approximate concave function to describe the SOC when recharging in both linear and non-linear regimes. Due to a decreasing charging rate after reaching 70%–80% battery capacity, most studies on EV-VRP/DRT problem modeling assume an 80% charge policy with a constant charging rate, although some recent studies consider non-linear charging approximation functions to model a more precise relationship between charging time and the amount of charged energy (Montoya et al., 2017; Kancharla and Ramadurai, 2020).

b. Charging policy

Charging policies can be classified into two categories: full charging and partial charging policies (Kucukoglu et al., 2021). The full charging policy assumes that EVs get recharged to their maximum battery capacity or a pre-defined SOC (e.g., 80%). Keskin and Çatay (2016) point out that the full charging policy is not realistic, in particular when the charging operation is close to the end of service time and vehicles do not need to be fully recharged to return to their depots. The partial charging policy reduces charging times and costs to meet customer demand (Keskin and Çatay, 2016; Pantelidis et al., 2021), and another approach involves battery-swapping technology that allows EVs to exchange their depleted batteries for fully charged ones to lower EV charging times (Vallera et al., 2021).

c. Charging cost

Most studies consider charging cost to be a linear function of the charged amount of energy, with a constant electricity price (Schneider et al., 2014). Some studies take into account service access and parking costs (Wang et al., 2018) or opportunity costs when vehicles are idle for recharging (Ma, 2021). As energy prices are variable depending on the time of the day, some recent studies incorporate time-dependent energy prices to reflect realistic charging costs (Sassi et al., 2014; Lin et al., 2021). Chen et al. (2016b) consider the heterogeneity of energy prices at different charging stations to more realistically reflect the ecosystem of charging service providers and their business models. Fehn et al. (2019) integrate dynamic electricity pricing to minimize the total charging costs of an e-fleet. They show that integrating time-dependent energy prices into EV-DRT system charging operations could lead to significantly lower charging costs.

d. Waiting time at charging stations

Most studies assume that EVs start recharging as soon as they arrive at public charging stations—the waiting time and number of available chargers at charging stations are not considered. Early studies consider charging station occupancy as a random variable to model the uncertain availability of chargers (Sweda et al., 2017). Keskin et al. (2019) model EV waiting times at charging stations as a queue with Poisson-distributed EV arrivals. The authors summarize the characteristics of EV-VRPs in terms of electricity consumption, recharging function (linear/non-linear), fleet composition (homogeneous/heterogeneous), objective functions, charger type (single/multiple), charging policy (full/partial recharge), and solution methods. Ammous et al. (2019) model EV waiting times at charging stations based on a multi-server queuing system for customer and vehicle arrivals. Kullman et al. (2018) consider that EVs can be charged either at depots (immediate charge) or public charging stations (based on a queuing system approach). Schoenberg and Dressler (2021) propose a multi-criteria EV route-planning method considering realistic EV energy consumption and queuing at charging stations. Waiting times at charging stations are estimated based on a centralized charging station database to collect EV users' intended charging stations in advance. Ma and Xie (2020) propose an online vehicle charger assignment model to allocate vehicles to chargers and minimize total charging operation time in terms of the access time, waiting time, and charging time of the fleet. Charging station occupancy information is communicated with the dispatch center for efficient vehicle–charger assignment policy development. However, the charging needs of other private/commercial EVs are not considered, which could impact vehicle waiting times when arriving at charging stations.

- Tactical decisions: This decision problem considers the optimization of fleet size and configuration over a medium-term horizon (weekly or monthly decision). The problem is to determine an optimal fleet size and configuration considering purchase and maintenance costs so as to satisfy a certain level of customer demand under the structural constraints of strategic decisions. The latter plays an essential role to ensure sufficient fleet size to meet customer demand in the medium-term and affects the level of service and operational costs at the operational level.
- Strategical decisions: This decision problem considers charging infrastructure planning to determine charging capacity and facility location and satisfy EV charging demands for a long-term horizon (annually). Strategic decisions influence charging time, waiting time, and the charging efficiency of the fleet. Strategic decisions require considering future transport market trends, economic perspectives, regulation and transport policy, new technologies, and customer demand forecasting, etc. This level of decision-making influences the decisions made at the tactical and operational levels.

## 3. State-of-the-art methodologies for EV-DRT system planning and management

In this section, we review and classify mathematical models and solution algorithms for the planning and management of EV-DRT systems. To limit the scope, we focus on three types of decision problems: charging scheduling, fleet size, and charging infrastructure. For reasons of space, we limit ourselves to key research studies and recent developments in these areas.

### *3.1. Charging scheduling (operational decisions)*

In recent decades, EV-DRT charging scheduling problems have been widely studied as an extension of classical VRPs by considering the limited driving range of EVs and the need to recharge at intermediate charging stations. Basic deterministic EV-DRT problems consider a fleet of homogeneous EVs to provide on-demand passenger transport services with time windows and vehicle capacity constraints. The objective is to minimize overall operational costs and customer inconvenience (Keskin et al., 2019; Schiffer et al., 2019; Asghari et al., 2021). This problem is generally formulated as mixed-integer linear programming (MILP) and can be solved using modern desktop computers to optimality for up to hundreds of customers within a couple of hours (Bongiovanni et al., 2019; Malheiros et al., 2020). For larger instances, various heuristics have been developed as well, including large neighborhood search (Schneider et al, 2015; Keskin et al., 2019), matheuristics (Malheiros et al., 2020), and various local search-based algorithms (see the recent review by Ho et al., 2018). While deterministic EV-VRP/DRT problems have been widely studied in the past, stochastic and dynamic EV-DRT problems are more challenging and have recently received increasing attention (Bongiovanni et al., 2019; Zalesak and Samaranayake, 2021). For deterministic problems, the reader is referred to the recent reviews by Asghari et al. (2021), Kucukoglu et al. (2021), and Xiao et al. (2021).

For dynamic EV-DRT problems, additional complexity for charging scheduling under uncertainty (e.g., stochastic customer arrivals, charging station availability, energy consumption, and traffic conditions, etc.) need to be considered. As new requests arrive on short notice or in real time, heuristics need to be developed to re-optimize the existing routes of vehicles to minimize additional costs for charging and serving customers. Berbeglia et al. (2010) summarize earlier studies of dynamic dial-a-ride problems. A review of more recent dial-a-ride problem (DARP) studies can be found in Ho et al. (2018). State-of-the-art methodologies can be lumped into two main categories: constrained optimization and approximate dynamic programming/reinforcement learning for optimal vehicle dispatching and charging scheduling. In this section, we focus on dynamic EV-DARP problems and discuss various recently proposed modeling approaches and solutions.

a. **Constrained optimization**: This approach formulates EV sequential dispatching and charging optimization problems as mixed-integer optimization problems and proposes heuristics to find approximate solutions for online applications. The optimization problems for request assignment and charging scheduling are considered separately and solved according to their respective objectives. For example, Iacobucci et al. (2019) propose a model predictive control approach for modeling shared autonomous electric vehicle charging scheduling optimization by integrating dynamic energy prices to minimize the overall charging cost of the fleet. Vehicles can sell back any remaining energy to the power grid by vehicle-to-grid (V2G) technologies. The proposed approach is tested using simulations on a small area (25 $km^2$) in Tokyo. Bongiovanni (2020) proposes an insertion heuristic to insert a new request into vehicles' routes with the lowest cost possible, considering both operational costs and customer inconvenience. A two-phase metaheuristic is proposed to find good solutions efficiently. Ma (2021) proposes two-stage charging scheduling and online vehicle-charger assignment to solve a dynamic EV-DARP using public charging stations. The author first derives a day-ahead charging plan (when and how much energy to charge over one day) as a battery replenishment optimization problem for each EV, based on their historical driving patterns, expected waiting times at charging stations, and variable electricity prices. Then, online vehicle-charger assignment optimization is solved for each decision epoch based on charging station occupancy states, assumed to be known by the vehicle dispatch center. The simulation studies show that the proposed method could significantly reduce charging costs while satisfying passenger demand. Zalesak and Samaranayake (2021) consider a dynamic EV-based ride-pooling problem using public charging stations. The authors propose a batch assignment optimization approach based on the sharability of ride requests in order to minimize operational costs while satisfying customer demand. A sequential optimization framework is proposed to optimize EV charging scheduling based on a two-stage approach (time scheduling, which involves determining when to charge over a longer planning horizon (e.g., $\geq$ 45 min.), and location scheduling—where to charge—over a short planning horizon (e.g., 15 min.)) with uncertain public charging station availability. The charging scheduling problems are formulated as two MILP problems to minimize shortages of vehicles and penalize vehicles with an insufficient SOC. The proposed methods are solved by heuristics and tested using Manhattan taxi-ride data. Yi and Smart (2021) propose an optimization model for jointly optimizing idle vehicle repositioning and charging decisions. However, this approach does not optimize the charging levels of EVs. Different from previous studies, recent studies consider the joint optimization of EV repositioning and partial recharge for online car-share rebalancing policy design (Ma et al., 2019; Pantelidis et al., 2021). The problem is modeled as p-median relocation based on a node-charge graph to jointly optimize EV repositioning and partial recharge decisions so as to minimize overall operational costs while satisfying customer demand.

b. **Approximate dynamic programming/reinforcement learning**: This approach considers EV dispatching and charging scheduling optimization based on the Markov decision process to model sequential decision-making under uncertainty. To solve the curse of dimensionality issue of Bellman's equation, different solution techniques drawn from approximate dynamic programming or reinforcement learning have been proposed. For example, Al-Kanj et al. (2020) propose an approximate dynamic programming approach to solve charging scheduling and vehicle dispatching for dynamic ride-hailing services. Each vehicle is modeled as an agent that makes sequential decisions among three possible actions: pass (continue being idle or in-service), charge (recharging battery), or assign a new customer. The parameterized reward functions are problem-specific and need to be fine-tuned. Vehicle–passenger assignment is solved in a centralized way considering all available vehicles, as a matching problem to maximize an action-value function. The proposed method is tested on numerical instances and obtains promising results. Shi et al. (2020) propose a reinforcement learning approach for dynamic EV-DARPs. EVs are modeled as agents to learn their state-action value functions based on a coarse space-time discretization. A feed-forward neural network approach is applied to obtain agents' approximated state-action value functions. Like many other reinforcement learning applications for ride-hailing operation optimization (Lin et al., 2018; Guo and Xu 2020), a centralized controller solves a linear request-assignment problem periodically. The objective is to minimize overall operational costs, charging costs, and customer waiting times. However, vehicle repositioning is not considered in this study. Kullman et al. (2021a) propose a deep reinforcement learning approach for dynamic EV-DARPs. Different from previous studies (Shi et al., 2020), which are based on a coarse space-time discretization of the action space, this study relaxes these limits by developing deep reinforcement learning to learn continuous state-action approximations. A simulation case study using Manhattan taxi-ride data from 2018 shows that the proposed method significantly outperforms two reference policies. However, the partial recharge policy is not considered. Different from the aforementioned study, Yu et al. (2021) propose an asynchronous learning approach to approximate the value function by sampling from future uncertain states for ride-hailing services using autonomous electric vehicles. The numerical study is based on trip data from the city of Haikou in China.

Table 1 summarizes the main characteristics of recent studies of dynamic EV-DARPs in terms of charging policies and methodologies for operational policy optimization.

Table 1. Summary of the dynamic EV-DRT literature.

| Reference | Problem | Charging | Methodology |
|---|---|---|---|
| Al-Kanj et al. (2020) | EAV-DARP | Linear, CP=A | Dynamic vehicle dispatch, repositioning, and recharge. Uses a hierarchical aggregation approach for value-function approximation under approximate dynamic programming. |
| Bongiovanni (2020) | EAV-DARP | Linear, partial recharge, CP=A | A two-stage heuristic approach. Uses a greedy insertion algorithm to insert new feasible requests and then re-optimize the decisions based on large neighborhood search heuristics. New recharging/idling decisions are checked at the end of vehicle routes. |
| Shi et al. (2020) | EAV-DARP | Linear, CP=A | Reinforcement learning approach to optimize vehicle routing and charging decisions under a spatio-temporal discretization framework. |
| Kullman et al., (2021a) | EV-DARP | Linear, full recharge, CP=B | Deep reinforcement learning to learn optimal routing and charging decisions under uncertainty. |
| Ma (2021) | EV-DARP | Linear, partial recharge, CP=C | Two-stage optimization approach. Determines when and how much energy to charge in the first stage and then where to charge in the second stage, based on charging station occupancy information. Mixed-integer optimization formulation. |
| Yu et al. (2021) | EAV-DARP | Linear, CP=B | Approximate dynamic programming under a Markov decision process. |
| Zalesak and Samaranayake (2021) | EV-DARP | Concave, CP= C | Two-stage optimization. Assigns new requests under the current charging schedule in the first stage, then optimizes the charging schedule (when and where to charge) given assigned requests. Mixed-integer optimization formulation. Charging priority depends on the sorted SOC of vehicles. |

Note. EAV: electric autonomous vehicle. DARP includes ride-pooling, ride-hailing, and DRT services. CP (charging policy): **A**. When an EV's battery level is lower than a threshold (around 10%), assign vehicles to nearby available charging stations to charge to a pre-defined maximum amount (around 80% of battery capacity); **B**. when and where to charge is determined by a sequential decision learning process, and a full-recharge (around 80%) policy is applied; **C**. charging amount and charging station assignment are based on an optimization model or heuristic to minimize overall charging operational costs or negative impacts on the service.

### *3.2. Fleet size and configuration (tactical decisions)*

The fleet-size problem is related to determining the capacity of a transport service so as to meet customer demand. It involves the trade-off between the investment costs and revenue losses when demand cannot be satisfied (Beaujon and Turnquist, 1991). The investment decision needs to jointly optimize the utilization of available vehicles and the fleet size under uncertain demand. In the context of an electrified transportation system, the fleet-size decision also requires considering the charging infrastructure's capacity to optimize charging operational costs of EVs. Existing studies of fleet-size planning can be classified into three categories according to how the trade-offs between customer demand, fleet acquisition costs, and vehicle charging demands are dealt with.

a. **Vehicle routing-based models with fixed demand**: This modeling approach extends static VRP modeling by integrating vehicle acquisition costs in the objective function so as to minimize the overall system costs required to serve given customer demand. The problems are generally formulated as mixed-integer optimization problems and are solved by state-of-the-art integer programming techniques for small instances and (meta)heuristics for large-scale instances. Charging scheduling subproblems are integrated into these models to track EV energy consumption and recharging so as to minimize fleet charging times/costs. For example, Hiermann et al. (2016) propose a joint optimization model for EV fleet configuration and routing. An adaptive large neighbourhood search (ALNS) is proposed to solve large-scale test instances. Rezgui et al. (2019) consider the joint fleet-size and routing optimization problem for modular electric vehicles. A variable neighborhood descent heuristic is proposed for larger extended Solomon's instances with up to 400 customers and hundreds of vehicles.
b. **Location routing models**: This approach considers the joint optimization of charging station location, fleet size, and vehicle routing decisions to meet deterministic customer demands. It involves the extension of VRP models by integrating strategic decisions (charging station locations) to achieve higher synergy and minimize the overall system operation cost from a long-term perspective. For example, Zhang et al. (2019) propose an optimization model for joint fleet-size and charging-infrastructure planning of autonomous electric vehicles. The objective function considers the annual investment costs of vehicles and charging facilities, operational costs (time and electricity), and maintenance costs to meet OD demands. Schiffer et al. (2019) provide a comprehensive review of location routing problems with intermediate stops.
c. **Simulation-based models under uncertain demand**: This approach applies simulation approaches to minimize the fleet-size requirements of on-demand mobility systems in order to meet specific demand scenarios. Using the simulation approach allows considering dynamic and stochastic environments and more realistic case studies. For example, Winter et al. (2018) propose a modeling framework to evaluate the performance and fleet requirements of automated DRT under different demand scenarios. The objective is to minimize the overall operational cost and passengers' generalized travel costs. Chen et al. (2016a) propose a multi-agent simulation approach for charging-infrastructure and fleet-size planning under different scenarios for autonomous electric vehicles.

d. **Network flow models**: This approach considers supply–demand interactions under a network flow modeling framework. Customer demand is expressed as flows in space (OD demand matrices) and time (discretized periods) for which the operator optimizes their fleet size, vehicle dispatches, and idle vehicle relocations so as to minimize the overall cost and unmet customer demand. Beaujon and Turnquist (1991) apply this approach for the joint optimization of fleet size and idle-vehicle allocation under stochastic demand.
e. **Stochastic/robust optimization models**: This approach aims to integrate different sources of uncertainty in fleet-size planning. Sayarshad and Tavakkoli-Moghaddam (2010) propose a stochastic optimization approach for multi-period rail-car fleet-size planning. The objective is to maximize the revenue generated by sending loaded cars between origins and destinations (ODs) while considering the transportation costs and unmet demand costs between ODs. A simulated annealing heuristic is proposed to solve the two-stage stochastic optimization problem. Schiffer and Walther (2018) propose a robust optimization approach for the joint optimization of fleet size, charging station locations, and routing under uncertain demand scenarios. The authors propose a hybrid ALNS solution. Guo et al. (2021) propose a robust optimization approach for the fleet-size minimization of taxi-like services using autonomous vehicles under the worst budget scenario in the uncertainty set. Shehadeh et al. (2021) propose a two-stage mixed-integer stochastic optimization approach and a distributionally robust two-stage optimization approach for fleet-size optimization and allocation to last-mile service regions under uncertain demand. The objective is to minimize the waiting times and riding times of customers. The solution is based on the sample average approximation approach to solve the stochastic optimization approximately.

Table 2 summarizes problem characteristics, problem formulations, objective functions, and solution algorithms.

Table 2. Summary of the fleet-size and configuration literature.

| Study | Problem | Features | Problem formulation | Objective function | Solution algorithm |
|---|---|---|---|---|---|
| Beaujon and Turnquist (1991) | Fleet size and idle vehicle relocation using ICVs | Variable demand over a sequence of decision periods, OD vehicle-flow optimization to meet demand | Stochastic optimization under stochastic demand | Maximize the expected profit as the difference between the revenue and the overall costs (fleet investment, operational cost and unmet demand loss) | Frank–Wolfe algorithm embedded heuristic |
| Hiermann et al. (2016) | Mixed fleet-size problem for vehicle-routing problem with time windows | Fixed demand, deterministic model, full-recharge policy | Mixed-integer programming | Vehicle acquisition costs and the total distance traveled | Branch and price algorithms for small instances and ALNS for large instances |
| Winter et al. (2018) | Minimum fleet-size problem for ADRT | Fixed demand, scenario-based simulations | Simulation of ADRT vehicle dispatching and routing policy | Vehicle acquisition costs and routing costs | Iterative simulations to find the fleet size with minimal system costs |
| Schiffer and Walther (2018) | Joint optimization of fleet size, charging station installation, and vehicle routing | EV location-routing problem with time windows and a partial-recharge policy | Mixed-integer programming and robust optimization | Vehicle routing, fleet acquisition, and charging station installation costs | ALNS |
| Zhang et al (2019) | Joint optimization of fleet size and charging stations for AEVs | Fixed OD demand, transport network flow, simplified routing and relocating of AEVs | Mixed-integer programming | Annual expected investment and operational costs | Branch-and-bound |
| Guo et al. (2021) | Robust minimization, fleet size optimization of on-demand ride services | Door-to-door service, each vehicle can serve one customer, zone-based OD demand, routing problem is not considered | Two-stage robust optimization | Minimize the fleet size to satisfy customer demand under the worst scenarios | Cutting planes |
| Rezgui et al. (2019) | Joint fleet-size and routing problem for EMVs | EMV routing by considering vehicle acquisition costs | Mixed-integer programming | Considers vehicle acquisition costs | VNS |
| Shehadeh et al. (2021) | Fleet allocation of last-mile on-demand feeder service under uncertain demand | Given passenger train arrivals and the stops on last-mile vehicle routes, determine fleet allocation to meet random last-mile service demand | Two-stage stochastic/robust optimization | Minimize total waiting times and riding times of customers | Sample average approximation |

Note. AEV: autonomous electric vehicle; ICV: internal combustion engine vehicle; ADRT: automated demand-responsive transport service; EMV: electric modular vehicles; ALNS: adaptive large neighborhood search; VNS: variable neighborhood search.

### *3.3. Charging infrastructure location and configuration (strategical decisions)*

Charging infrastructure planning needs to consider the characteristics of charging technologies, procurement, installation and operational costs, location costs, user charging needs/usage scenarios, power grid supply and compatibility with the charging power of vehicles, etc. The decision of charging technologies influences the charging times of vehicles, which in turn affects the availability of both chargers and vehicles to serve customers. Different mathematical models have been proposed in the past to address either public or company-owned charging-station planning. Public charging planning considers the charging needs of private EVs (Liu and Wang, 2017) or e-taxis (Jung et al., 2014). The main modeling approaches can be classified into node-based facility location and path-based facility location approaches (Jung et al., 2014; Kchaou-Boujelben, 2021). The reader is referred to the recent reviews of this topic (Rahman et al., 2016; Deb et al., 2018; Pagany et al., 2019; Kchaou-Boujelben et al., 2021). In the current paper, we focus on the mathematical modeling and solution algorithms for charging infrastructure planning involving the vehicle-routing decisions of EV-DRT systems.

a. **Set covering**: Kunith et al. (2014) propose a capacitated set-covering model to plan the number and locations of fast-charging stations for busline operations considering daily customer demand. Realistic energy consumption scenarios are considered to take into account the charging needs of bus operations under different scenarios. Wu et al. (2021) propose fast-charging facilities for an e-bus system at existing bus terminals to minimize the overall investment, the maintenance costs of charging facilities, the access costs for recharging, and the power-loss costs under charging load capacity constraints. An (2020) formulates a variant of the set-covering model to jointly optimize bus charging-station locations and fleet size by considering time-dependent energy prices and stochastic bus charging demand.

b. **Location-routing optimization**: This approach formulates the charging infrastructure planning problem as a joint optimization problem with vehicle-route and/or fleet-size planning. Based on the characteristics of demand, the proposed approaches can be further classified into classical integer programming and stochastic/robust optimization to address different sources of uncertainty. For example, Schiffer and Walther (2017) propose a mixed-integer location-routing model to jointly minimize the number of charging stations, the fleet size, and routing costs by considering partial recharge for VRP with time-window constraints. The proposed model is tested on the instances of Schneider et al. (2014), with up to 100 customers, and solved by commercial solvers. Hua et al. (2019) propose a multi-stage stochastic optimization problem for charging-infrastructure planning of an electric car-sharing system. The model takes into account the joint optimization of long-term charging infrastructure planning and short-term vehicle relocation and charging operations of the fleet under uncertain multi-period demands. Random customer demand is modeled using the scenario tree approach. The objective is to minimize the overall system cost over multiple planning periods. Stumpe et al. (2021) propose a mixed-integer programing model for charging infrastructure locations and electric busline operation optimization for a set of bus trips. The authors propose a sensitivity analysis approach to identify persistent structures of the solutions given uncertain input parameter distributions and configurations based on realistic electric bus operation data.

c. **Bi-level optimization-simulation approach**: This approach considers a bi-level modeling structure by iteratively optimizing the location and number of charging stations at the upper level while simulating charging operations at the lower level to obtain the system performance in terms of charging operational delays or vehicle idle times. This approach can flexibly take into account different sources of uncertainty and explicitly considers charging waiting times based on EV arrival and charging service rates at charging stations. For example, Jung et al. (2014) propose a bi-level optimization-simulation approach to locate e-taxi charging stations in an urban area. The upper-level problem is modeled as a multiple server location problem under the number of charger installations at each candidate charging location. Ma and Xie (2020) propose a bi-level optimization-simulation approach for charging-infrastructure planning for electric microtransit systems. The authors consider the sub-problem of online vehicle–charger assignment optimization to minimize the idle times of vehicles when recharging. A surrogate-based optimization approach is proposed for its application in a realistic simulation case study in Luxembourg. Lokhandwala and Cai (2020) propose an agent-based simulation approach for charging-infrastructure planning under different demand scenarios. The model allows considering the number of charging stations and plugs, as well as the problem of extensions of new charging infrastructure. The modeling framework consists of generating customer demand, based on which EV charging demand is explicitly simulated with queuing dependent on EV arrival and service rates.

Table 3 summarizes recent developments in charging station location planning for electrified transportation systems.

Table 3. Summary of charging infrastructure planning literature.

| Study | Problem | Features | Problem formulation | Objective function | Solution algorithm |
|---|---|---|---|---|---|
| Kunith et al. (2014) | Fast-charging infrastructure configuration for an electric busline | Realistic energy consumption, operational and technical constraint | Mixed-integer programming | Minimize the overall charging infrastructure investment costs under the constraints of serving daily customer demand | Commercial solver |
| | | modeling for electric buses, set-covering formulation | | | |
| Jung et al. (2014) | Charging-infrastructure planning for an electric taxi fleet in an urban area | Multiple server allocation for modeling queuing delays at charging stations | Bi-level optimization simulation | Minimize overall EV access times and waiting times at charging stations | Genetic algorithm |
| Schiffer and Walther (2017) | Charging-infrastructure, fleet-size, and routing optimization of EVRP | Location-routing planning, partial recharge, and time window constraints | Mixed-integer programming | Minimize overall system costs including charging infrastructure, fleet acquisition, and routing costs | Commercial solver |
| Hua et al. 2019 | Joint charging-infrastructure and vehicle-operation optimization for electric car-sharing system | Multiple decision periods with uncertain demand based on the scenario tree approach | Multi-stage stochastic optimization | Minimize the overall expected system costs over multiple planning periods | Projected subgradient algorithm |
| Ma and Xie (2020) | Fast-charging station location optimization for microtransit service | Simulation approach with queuing at charging stations | Bi-level optimization simulation | Minimize the fleet's total idle time for charging operations while considering optimal vehicle–charging station assignment to minimize vehicle idle times | Surrogate-assisted optimization algorithm |
| Lokhandwala and Cai (2020) | Charging-infrastructure, fleet-size, and routing optimization of EVRP shared AEVs | Agent-based simulation, queuing at charging stations is based on the arrival and service rates of AEVs | Bi-level optimization simulation | Minimize total waiting times at charging stations over the planning horizon when the number of charging stations and plugs is constrained by a fixed budget | Genetic algorithm |
| Lin et al. (2019) | Charging-infrastructure location and configuration planning with integrated power-grid impact to minimize power loss | Multistage planning problems with interplays between the transport system and power grid | Multistage | Minimize the overall system costs including charging station investment, energy consumption costs, and power grid extension costs | Commercial solver |
| Stumpe et al. (2021) | Joint optimization of charging infrastructure and busline operations | Departure-time-based service trips, sensitivity analysis for different input parameter uncertainties | Mixed-integer programming | Minimize the overall system costs of charging infrastructure, bus investment costs, personnel, and energy consumption costs | VNS |
| Wu et al. (2021) | Fast-charging station planning at electric bus terminals | Considers maximum charging load constraints at bus terminals | Simulation-optimization approach | Minimize overall charging station investment costs | Particle swarm optimization |
| An (2020) | Joint optimization of fleet-size and charging-station planning under | Considers stochastic charging demand, time-dependent energy price | Mixed-integer programming | Minimize the overall system cost including charging station acquisition, maintenance, and the access costs of vehicles | Lagrangian relaxation |

| | | | | |
|---|---|---|---|---|
| | stochastic bus-charging demand | | | |

Note. AEV: autonomous electric vehicle; VNS: variable neighborhood search; EVRP: electric vehicle routing problems.

## 4. Open-access datasets, test instances, and software

In this section, we summarize the available open-access online resources including trips datasets, numerical test instances, and software for EV-VRP and its variants (see Table 4). These freely available datasets and software could be adapted to generate new test instances with new algorithms. Two datasets by Schneider et al. (2014) for electric vehicle routing problems with time windows (E-VRPTW), extended from the benchmark instances of Solomon (1987), have been widely used to test different variants of EV-VRP. Felipe et al. (2014) provide large instances with 100, 200, and 400 randomly distributed customers. Mendoza et al. (2014) collect several VRP- and EV-VRP-related test instances. Bongiovanni (2020) provides two sets of small instances with up to 50 customers based on randomly generated customers and Uber ride data. Some authors use freely available ride data from ride-hailing companies to test the performance of proposed operational policies for dynamic ride-hailing systems (Kullman et al., 2021a; Yu et al., 2021). Froger et al. (2021) provide 120 test instances for EV-VRP with non-linear charging functions and capacitated charging stations. The exact up-to-date and heuristic solutions for EV-VRPTW are provided in Kucukoglu et al. (2021).

In terms of solution algorithms and software, most studies utilize commercial solvers like CPLEX and Gurobi to solve the mixed-integer programming problems and obtain exact solutions. There are few freely available codes for the heuristics shared among the scientific community. An exception can be found in Kullman et al. (2021b), who publish their Python package for solving the exact fixed-route vehicle-charging problem using the labeling algorithm. The recently developed general VRP solver has good potential for application in solving exact EV-VRP variants (Pessoa et al., 2020).

Table 4. Open-access datasets, test instances, and software.

| Reference | Feature | # of instances | Link |
|---|---|---|---|
| ***Datasets/test instances/solutions*** | | | |
| Schneider et al. (2014) | E-VRPTW, widely used datasets for different variants of VRP and location-routing problems using EVs | A set of small instances (5, 10, and 15 customers ) and a set of large instances (up to 100 customers and 21 charging stations) | https://data.mendeley.com/datasets/h3mrm5dhxw/1 |
| Mendoza et al. (2014) | Vehicle-routing problem repository (VRP-REP) | | http://www.vrp-rep.org/ |
| Felipe et al. (2014) | Green-vehicle routing problem with multiple technologies and partial recharges | 3 sets of 20 instances with 100, 200, and 400 customers | http://www.mat.ucm.es/~gregoriotd/GVRPen.htm |
| Bongiovanni (2020) | Uber ride-share data from San Francisco | | https://github.com/dima42/uber-gps-analysis/tree/master/gpsdata |
| Bongiovanni (2020) | EAV-DARP instances | Two datasets adapted from Cordeau (2006) and the Uber dataset mentioned above; each contains 14 instances (up to 5 vehicles and 50 customers) and solutions for the Uber test dataset | https://luts.epfl.ch/wpcontent/uploads/2019/03/e_ADARP_archive.zip |
| Kullman et al. (2021a) | Ride-hailing data from New York City | | https://www1.nyc.gov/site/tlc/about/tlc-trip-record-data.page |
| Yu et al. (2021) | Trip data from Didi Chuxing ride-hailing services | | https://outreach.didichuxing.com/research/opendata/ |
| Froger et al. (2021) | Test instances for EVRP with non-linear charging functions and capacitated stations | 120 instances from Montoya et al. (2017) | https://www.math.u-bordeaux.fr/~afroger001/research.html |
| Kucukoglu et al. (2021) | Summary of the best exact and heuristic solutions to date for E-VRPTW | | |
| ***Software*** | | | |
| Kullman et al. (2021b) | Python package | Exact labeling algorithm to solve the fixed-route vehicle-charging problem | https://www.math.u-bordeaux.fr/~afroger001/research.html |
| Pessoa et al. (2020) | Open-access solver developed by a research group at the University of Bordeaux | Branch-cut-and-price-based exact solver for VRP-related problems, interface implemented in Julia and JuMP package | https://vrpsolver.math.u-bordeaux.fr/ |

Note. Format (a–b/c): number of instances – number of customers / number of charging stations.

## 5. Conclusions and future research directions

The accelerated electrification of transport systems with EVs has brought new challenges for charging scheduling, fleet management, and charging infrastructure location and configuration planning. We have summarized recent developments and mathematical modeling approaches and identified future research directions for strategic, tactical, and operational decisions for EV-DRT systems. Moreover, existing open-access datasets, numerical test instances, and software are listed for future research in EV-DRT and related problems. Future research directions are discussed as follows.

### *System-level integration with the power grid*

- Integrating impact on power grids: While significant research has contributed to EV charging scheduling in the context of urban logistics or DRT, few studies integrate the impact of fleet charging on the power grid. Future extensions could address this by considering the impact on the power distribution network according to the number of charging plugs and space-time power-supply constraints (Mohamed et al., 2017; Abdelwahed et al., 2020). Moreover, studying smart dynamic fleet charging/discharging strategies—including V2G technology—to enhance grid stability and increase TNC revenue is a promising research direction.
- Integrating time-dependent energy prices: Integrating time-dependent energy prices to minimize TNC fleet-charging management could significantly reduce daily charging costs (Lin et al., 2021). However, most existing studies ignore this aspect by assuming unitary energy prices. Future extensions could develop new charging strategies by considering time-varying energy prices and evaluating the impact on cost and other system-performance metrics.
- Integrating smart grids for fleet charging management: EV charging management can further consider frequency regulation support of the power grid as an energy storage device. With V2G technology, EVs can further gain revenue by applying smart charging/discharging strategies during off-peak/peak hours (Hu et al., 2016). The cooperative game-theoretical approach (Ziad et al., 2019) provides a methodological framework to model the interactions of EV fleets and different actors in the energy market (aggregators, distribution system operators, energy prosumers/producers, etc).

- Increasing the use of renewable energy sources for charging: To further minimize impacts in terms of climate change, the use of renewable energy for EVs could be enhanced by integrating the modeling of energy-demand interactions. Wellik et al. (2021) propose a simulation approach to jointly model the interactions of a grid operator and transit bus with V2G technology charging operations so as to minimize the energy supply costs of the grid operator and the total charging costs of e-buses by ensuring a greater use of renewable energy sources for charging. Future research could investigate dynamic pricing mechanisms using a game-theoretical approach to incentivize the participation of bus operators in grid-support services.

***Multi-period planning and decision support-system development***

- Integrate multi-period planning: Significant research has focused on the single-period problem for EV routing and charging management planning with given demands. Further research could study joint strategic, tactical, and operational decision planning under a multi-period planning horizon. Under this modeling framework, demand uncertainty and dynamic resource assignment could be integrated over longer planning periods, for which stochastic or robust optimization methods could be developed under different scenarios (Sayarshad and Tavakkoli-Moghaddam, 2010; Bertsimas et al., 2017).
- Developing a decision support system for EV-DRT system planning and policy evaluation: Developing such a decision support system would help transport operators and policymakers evaluate and test their business models before and during the deployment of their services. For policymakers, such tools enable evaluating the impacts of different policies on social welfare (Balac et al., 2019). Existing studies are mainly based on combustion-engine vehicles (Horn, 2002; Dias et al., 2012). Using a systems engineering approach and enhancing collaboration between operators and stakeholders could help develop an impactful decision support system to promote EV-DRT system deployment in the future (Danandeh et al., 2016; Xie et al., 2016).

***Dynamic charging scheduling under uncertainty***

- Developing advanced deep reinforcement-learning techniques for addressing state-action value approximation involving continuous state/action spaces: The recent study by Kullman et al. (2021) shows promising results compared to a re-optimization policy for ride-hailing services. Future research could extend this study to other on-demand mobility services and compare its results with other optimization-based approaches (Zalesak and Samaranayake, 2021).
- Multimodal integration with EV-DRT: Several studies have proposed a reinforcement-learning-based approach to optimize the dispatching and routing of ride-hailing/ride-sharing services using EVs under a stochastic environment. However, these studies do not consider service integration with the transit system. It is a natural extension to integrate transit systems into these on-demand mobility systems to provide users with seamless multimodal solutions to future requests based on the reinforcement-learning approach. Moreover, it is desirable to make a set of numerical test instances freely available to support algorithm design and comparison.

***Modeling challenges with new technologies***

- Developing new charging strategies involving V2G technology: V2G technology provides EV fleets with the capability of providing grid services to gain revenue from smart charging strategies. Existing studies mainly focus on the charging control strategy of private EVs to encourage their participation in grid services (Saad et al., 2012). Further research could study smart charging strategies using V2G technology to maximize the revenue from charging/discharging operations and provide grid services to reduce their impact on the grid.
- Develop incentives to encourage TNC participation in grid-support services: Given the expected large-scale adoption of EVs in the near future, their impact on the grid needs to be assessed and remedial solutions using EV energy storage and discharge capabilities via V2G technology need to be developed. Existing studies mainly focus on private EV charging strategy development, and investigations of electric vehicle on-demand mobility services are few. The game-theoretical approach provides a theoretical framework to model the interactions between different actors (e.g., energy producers, grid operators, prosumers, EVs, and other customers). It allows designing efficient control and pricing policies by considering the competition and cooperative behaviors of different agents (Saad et al., 2012; Zhu et al., 2018; Iacobucci et al.,2019; Zhang et al.,2021). Future research could develop dynamic pricing mechanisms to allow fleet operators to lower their energy consumption costs, increase the use of renewables, and support frequency-regulation services for the power grid.

References

1. Abdelwahed, A., van den Berg, P.L., Brandt, T., Collins, J., Ketter, W., (2020). Evaluating and optimizing opportunity fast-charging schedules in transit battery electric bus networks. Transp. Sci. 54, 1601–1615.
2. Al-Kanj L., Nascimento J., Powell W. B., (2020). Approximate dynamic programming for planning a ride-hailing system using autonomous fleets of electric vehicles. Eur. J. Oper. Res., 284(3), 1088–1106.
3. Ammous, M.; Belakaria, S.; Sorour, S.; Abdel-Rahim, A. (2019). Optimal Cloud-Based Routing with In-Route Charging of Mobility-on-Demand Electric Vehicles. IEEE Trans. Intell. Transp. Syst., 20, 2510–2522.
1. An, K. (2020). Battery electric bus infrastructure planning under demand uncertainty. Transportation Research Part C: Emerging Technologies, 111, 572-587.
2. Asghari, M., Al-e, S. M. J. M. (2021). Green vehicle routing problem: A state-of-the-art review. Inter. J. Prod. Econo., 231, 107899.
3. Balac, M., Becker, H., Ciari, F., Axhausen, K.W., (2019). Modeling competing free-floating carsharing operators – A case study for Zurich, Switzerland. Transp. Res. Part C Emerg. Technol. 98, 101–117.
4. Beaujon, G. J., Turnquist, M. A. (1991). A model for fleet sizing and vehicle allocation. Transp. Sci., 25(1), 19-45.
5. Berbeglia, G., Cordeau, J.F., Laporte, G., (2010). Dynamic pickup and delivery problems. Eur. J. Oper. Res. 202 (1), 8–15.
4. Bertsimas, D., Griffith, J.D., Gupta, V., Kochenderfer, M.J., Mišić, V. V., (2017). A comparison of Monte Carlo tree search and rolling horizon optimization for large-scale dynamic resource allocation problems. Eur. J. Oper. Res. 263, 664–678.

5. Bongiovanni, C., Kaspi, M., Geroliminis, N. (2019). The electric autonomous dial-a-ride problem. Transp. Res. Part B: Methodol., 122, 436-456.
6. Bongiovanni, C., (2020). The electric autonomous dial-a-ride problem. PhD dissertation. EPFL.
7. Brandstätter, G., Leitner, M., Ljubić, I. (2020). Location of charging stations in electric car sharing systems. Transp. Sci., 54(5), 1408-1438.
8. Chen, T. D., Kockelman, K. M., Hanna, J. P. (2016a). Operations of a shared, autonomous, electric vehicle fleet: Implications of vehicle & charging infrastructure decisions. Transp. Res. Part A: Policy and Practice, 94, 243-254.
6. Chen, T., Zhang, B., Pourbabak, H., Kavousi-Fard, A., Su, W. (2016b). Optimal routing and charging of an electric vehicle fleet for high-efficiency dynamic transit systems. IEEE Trans. on Smart Grid, 9(4), 3563-3572.
7. Chow, J. Y., Rath, S., Yoon, G., Scalise, P., Saenz, S. A. (2020). Spectrum of Public Transit Operations: From Fixed Route to Microtransit.
8. Cordeau, J. F. (2006). A Branch-and-Cut Algorithm for the Dial-a-Ride Problem. Oper. Res. 54(3), 573–586.
9. Danandeh, A., Zeng, B., Caldwell, B., Buckley, B. (2016). A decision support system for fuel supply chain design at tampa electric company. INFORMS J. Applied Analytics, 46(6), 503-521.
10. Deb, S., Tammi, K., Kalita, K., Mahanta, P. (2018). Review of recent trends in charging infrastructure planning for electric vehicles. Wiley Interdisciplinary Reviews: Energy and Environment, 7(6), e306.
11. Deng, R., Liu, Y., Chen, W., & Liang, H. (2019). A Survey on Electric Buses—Energy Storage, Power Management, and Charging Scheduling. IEEE Trans.on Intell. Transp. Sys., 22(1), 9-22.
12. Dias, A., Telhada, J., Carvalho, M.S., (2012). Simulation approach for an integrated decision support system for demand responsive transport planning and operation. In: 10th Int. Ind. Simul. Conf. 2012, ISC 2012 130–138.
13. EEA (2021) Greenhouse gas emissions from transport in Europe. https://www.eea.europa.eu/data-and-maps/indicators/transport-emissions-of-greenhouse-gases-7/assessment.
14. Fehn, F., Noack, F., Busch, F. (2019). Modeling of mobility on-demand fleet operations based on dynamic electricity pricing. In 2019 6th International Conference on Models and Technologies for Intelligent Transportation Systems (MT-ITS) (pp. 1-6). IEEE.
15. Felipe, Á., Ortuño, M. T., Righini, G., Tirado, G. (2014). A heuristic approach for the green vehicle routing problem with multiple technologies and partial recharges. Transp. Res. Part E: Logistics and Transp. Rev., 71, 111-128.
16. Fiori, C., Ahn, K., Rakha, H. A. (2016). Power-based electric vehicle energy consumption model: Model development and validation. Applied Energy, 168, 257-268.
17. Froger, A., Jabali, O., Mendoza, J. E., Laporte, G. (2021). The electric vehicle routing problem with capacitated charging stations. Transp. Sci. (to appear).
18. Guo, G., Xu, Y. (2020). A deep reinforcement learning approach to ride-sharing vehicles dispatching in autonomous mobility-on-demand systems. IEEE Intell. Transp. Sys. Magazine.
19. Guo, Z., Hao, M., Yu, B., Yao, B. (2021). Robust minimum fleet problem for autonomous and human-driven vehicles in on-demand ride services considering mixed operation zones. Transp. Res. Part C: Emerging Technologies, 132, 103390.
20. Hazan J, Lang N, Wegscheider AK, Fassenot B, (2019). On-Demand Transit Can Unlock Urban Mobility.
21. Hiermann, G., Puchinger, J., Ropke, S., Hartl, R. F. (2016). The electric fleet size and mix vehicle routing problem with time windows and recharging stations. Eur. J. Oper. Res., 252(3), 995-1018.

22. Ho, S. C., Szeto, W. Y., Kuo, Y. H., Leung, J. M., Petering, M., Tou, T. W. (2018). A survey of dial-a-ride problems: Literature review and recent developments. Transp. Res. Part B: Methodological, 111, 395-421.
23. Horn, M.E.T., (2002). Multi-modal and demand-responsive passenger transport systems: A modelling framework with embedded control systems. Transp. Res. Part A Policy Pract. 36, 167–188.
24. Hu, J., Morais, H., Sousa, T., Lind, M., (2016). Electric vehicle fleet management in smart grids: A review of services, optimization and control aspects. Renew. Sustain. Energy Rev. 56, 1207–1226.
25. Hua, Y., Zhao, D., Wang, X., Li, X. (2019). Joint infrastructure planning and fleet management for one-way electric car sharing under time-varying uncertain demand. Transp. Res. Part B: Methodological, 128, 185-206.
26. Iacobucci, R., McLellan, B., Tezuka, T. (2019). Optimization of shared autonomous electric vehicles operations with charge scheduling and vehicle-to-grid. Transp. Res. Part C: Emerging Technologies, 100, 34-52.
27. Jenn A, (2019). Electrifying Ride-sharing: Transitioning to a Cleaner Future. UC Davis: National Center for Sustainable Transportation.
28. Jung, J., Chow, J. Y., Jayakrishnan, R., Park, J. Y. (2014). Stochastic dynamic itinerary interception refueling location problem with queue delay for electric taxi charging stations. Transp. Res. Part C: Emerging Technologies, 40, 123-142.
29. Kancharla, S. R., Ramadurai, G. (2020). Electric vehicle routing problem with non-linear charging and load-dependent discharging. Expert Sys. with Applications, 160, 113714.
30. Kchaou-Boujelben, M. (2021). Charging station location problem: A comprehensive review on models and solution approaches. Transp. Res. Part C: Emerging Technologies, 132, 103376.
31. Keskin, M., Çatay, B. (2016). Partial recharge strategies for the electric vehicle routing problem with time windows. Transp. res. Part C: emerging technologies, 65, 111-127.
32. Keskin, M., Laporte, G., Çatay, B. (2019). Electric Vehicle Routing Problem with Time-Dependent Waiting Times at Recharging Stations. Comput. Oper. Res, 107, 77–94.
33. Kucukoglu, I., Dewil, R., Cattrysse, D. (2021). The electric vehicle routing problem and its variations: A literature review. Computers & Industrial Engineering, 107650.
34. Kullman, N., Goodson, J., Mendoza, J. E. (2018). Dynamic electric vehicle routing: heuristics and dual bounds. https://hal.archives-ouvertes.fr/view/index/docid/1928730.
35. Kullman, N. D., Cousineau, M., Goodson, J. C., Mendoza, J. E. (2021a). Dynamic Ride-Hailing with Electric Vehicles. Transportation Science, https://doi.org/10.1287/trsc.2021.1042.
36. Kullman, N. D., Froger, A., Mendoza, J. E., & Goodson, J. C. (2021b). frvcpy: An Open-Source Solver for the Fixed Route Vehicle Charging Problem. INFORMS Journal on Computing, 33(4):1277-1283.
37. Kunith, A., Goehlich, D., Mendelevitch R. (2014). Planning and Optimization of a Fast Charging Infrastructure for Electric Urban Bus Systems. In: Proceedings of the 2nd International Conference on Traffic and Transport Engineering, pp. 43-51.
38. Lin, K., Zhao, R., Xu, Z., Zhou, J. (2018). Efficient large-scale fleet management via multi-agent deep reinforcement learning. In Proceedings of the 24th ACM SIGKDD International Conference on Knowledge Discovery & Data Mining (pp. 1774-1783).
39. Lin, Y., Zhang, K., Shen, Z. J. M., Ye, B., Miao, L. (2019). Multistage large-scale charging station planning for electric buses considering transportation network and power grid. Transportation Research Part C: Emerging Technologies, 107, 423-443.
40. Lin, B., Ghaddar, B., Nathwani, J. (2021). Electric vehicle routing with charging/discharging under time-variant electricity prices. Transp. Res. Part C: Emerging Technologies, 130, 103285.

41. Liu, H., Wang, D. Z. (2017). Locating multiple types of charging facilities for battery electric vehicles. Transp. Res. Part B: Methodological, 103, 30-55.
42. Lokhandwala, M., Cai, H. (2020). Siting charging stations for electric vehicle adoption in shared autonomous fleets. Transp. Res. Part D: Transport and Environment, 80, 102231.
43. Ma, T-Y., Rasulkhani, S., Chow, J. Y., Klein, S. (2019). A dynamic ridesharing dispatch and idle vehicle repositioning strategy with integrated transit transfers. Transp. Res. Part E: Logistics and Transportation Review, 128, 417-442.
44. Ma, T-Y, Pantelidis, T. P., Chow, J.Y.J. (2019b). Optimal queueing-based rebalancing for one-way electric carsharing systems with stochastic demand. In: Proceedings of the 98th Annual Meeting of the Transportation Research Board, Paper No. 19-05278, 2019, pp. 1–17.
45. Ma T-Y (2021) Two-stage battery recharge scheduling and vehicle-charger assignment policy for dynamic electric dial-a-ride services. PLoS ONE 16(5): e0251582.
46. Ma, T-Y, Xie, S (2021). Optimal fast charging station locations for electric ridesharing service with online vehicle-charging station assignment. Transp. Res. Part D: Transport and Environment, 90, 102682.
47. Malheiros, I., Ramalho, R., Passeti, B., Bulhões, T., Subramanian, A., (2020). A hybrid algorithm for the multi-depot heterogeneous dial-a-ride problem. Comput. Oper. Res. 105196.
48. Mohamed, M., Farag, H., El-Taweel, N., Ferguson, M., (2017). Simulation of electric buses on a full transit network: Operational feasibility and grid impact analysis. Electr. Power Syst. Res. 142, 163–175.
49. Montoya, A., Guéret, C., Mendoza, J. E., Villegas, J. G. (2017). The electric vehicle routing problem with nonlinear charging function. Transp. Res. Part B: Methodological, 103, 87-110.
50. Mendoza, JE, Guéret,C, Hoskins, M, Lobit, H, Pillac, V, Vidal, T, Vigo, D. (2014). VRP-REP: the vehicle routing community repository. Third meeting of the EURO Working Group on Vehicle Routing and Logistics Optimization (VeRoLog). Oslo, Norway.
51. Olsen, N. (2020). A literature overview on scheduling electric vehicles in public transport and location planning of the charging infrastructure. Discussion Papers 2020/16, Free University Berlin, School of Business & Economics.
52. Pagany, R., Ramirez Camargo, L., Dorner, W. (2019). A review of spatial localization methodologies for the electric vehicle charging infrastructure. International Journal of Sustainable Transportation, 13(6), 433-449.
53. Pantelidis, T., Li, L., Ma, T.-Y., Chow, J.Y.J., Jabari, S.E. (2021). Node-charge graph-based online carshare rebalancing with capacitated electric charging. Transp. Sci. https://doi.org/10.1287/trsc.2021.1058
54. Pavlenko AN, Slowik P, Lutsey N, (2019). When does electrifying shared mobility make economic sense? https://www.theicct.org/publications/shared-mobility-economic-sense.
55. Pessoa, A., Sadykov, R., Uchoa, E., Vanderbeck, F. (2020) A generic exact solver for vehicle routing and related problems. Mathematical Programming B, 183:483-523.
56. Rahman, I., Vasant, P. M., Singh, B. S. M., Abdullah-Al-Wadud, M., Adnan, N. (2016). Review of recent trends in optimization techniques for plug-in hybrid, and electric vehicle charging infrastructures. Renewable and Sustainable Energy Reviews, 58, 1039-1047.
57. Rezgui, D., Siala, J. C., Aggoune-Mtalaa, W., Bouziri, H. (2019). Application of a variable neighborhood search algorithm to a fleet size and mix vehicle routing problem with electric modular vehicles. Computers & Industrial Engineering, 130, 537-550.
58. Saad, W., Han, Z., Poor, H.V., Başar, T., (2012). Game-theoretic methods for the smart grid: An overview of microgrid systems, demand-side management, and smart grid communications. IEEE Signal Process. Mag. 29, 86–105.

59. Sassi, O., Cherif, W. R., Oulamara, A. (2014). Vehicle routing problem with mixed fleet of conventional and heterogeneous electric vehicles and time dependent charging costs. doi.org/10.5281/zenodo.1099756
60. Sayarshad, H. R., Tavakkoli-Moghaddam, R. (2010). Solving a multi periodic stochastic model of the rail–car fleet sizing by two-stage optimization formulation. Applied Mathematical Modelling, 34(5), 1164-1174.
61. Schoenberg, S., Dressler, F. (2021). Reducing Waiting Times at Charging Stations with Adaptive Electric Vehicle Route Planning. arXiv:2102.06503.
62. Schiffer, M., Walther, G. (2017). The electric location routing problem with time windows and partial recharging. Eur. J. Oper. Res., 260(3), 995-1013.
63. Schiffer, M., Walther, G. (2018). Strategic planning of electric logistics fleet networks: A robust location-routing approach. Omega, 80, 31-42.
64. Schiffer, M., Schneider, M., Walther, G., Laporte, G. (2019). Vehicle routing and location routing with intermediate stops: A review. Transp. Sci., 53(2), 319-343.
65. Schneider, M., Stenger, A., Goeke, D. (2014). The electric vehicle-routing problem with time windows and recharging stations. Transp. Sci., 48(4), 500-520.
66. Schneider, M., Stenger, A., Hof, J. (2015). An adaptive VNS algorithm for vehicle routing problems with intermediate stops. OR Spectrum, 37(2), 353-387.
67. Shehadeh, K. S., Wang, H., Zhang, P. (2021). Fleet sizing and allocation for on-demand last-mile transportation systems. Transp. Res. Part C: Emerging Technologies, 132, 103387.
68. Shen, Z. J. M., Feng, B., Mao, C., Ran, L. (2019). Optimization models for electric vehicle service operations: A literature review. Transp. Res. Part B: Methodological, 128, 462-477.
69. Shi, J., Gao, Y., Wang, W., Yu, N., Ioannou, P. A. (2020). Operating electric vehicle fleet for ride-hailing services with reinforcement learning. IEEE Transactions on Intelligent Transportation Systems, 21(11), 4822-4834.
70. Solomon, M. (1987) Algorithms for the Vehicle Routing and Scheduling Problem with time Window Constraints. Operations Research, 35, 254-265.
71. Sovacool, B.K., Kester, J., Noel, L., Zarazua de Rubens, G., (2020). Actors, business models, and innovation activity systems for vehicle-to-grid (V2G) technology: A comprehensive review. Renew. Sustain. Energy Rev. 131, 109963.
72. Spöttle M et al. (2018). Research for TRAN Committee – Charging infrastructure for electric road vehicles, European Parliament, Policy Department for Structural and Cohesion Policies, Brussels.
73. Stumpe, M., Rößler, D., Schryen, G., Kliewer, N. (2021). Study on sensitivity of electric bus systems under simultaneous optimization of charging infrastructure and vehicle schedules. EURO Journal on Transportation and Logistics, 10, 100049.
74. Sweda, T. M., Dolinskaya, I. S., Klabjan, D. (2017). Adaptive routing and recharging policies for electric vehicles. Transp. Sci., 51(4), 1326-1348.
75. Vallera, A. M., Nunes, P. M., & Brito, M. C. (2021). Why we need battery swapping technology. Energy Policy, 157, 112481.
76. Volinski, J. (2019) Microtransit or General Public Demand Response Transit Services: State of the Practice (2019)
77. Wang, Y., Bi, J., Guan, W., Zhao, X. (2018). Optimising route choices for the travelling and charging of battery electric vehicles by considering multiple objectives. Transp. Res. Part D: Transport and Environment, 64, 246-261.
78. Wellik, T.K., Griffin, J.R., Kockelman, K.M., Mohamed, M., (2021). Utility-transit nexus: Leveraging intelligently charged electrified transit to support a renewable energy grid. Renew. Sustain. Energy Rev. 139, 110657.

79. Winter, K., Cats, O., Correia, G., van Arem, B. (2018). Performance analysis and fleet requirements of automated demand-responsive transport systems as an urban public transport service. International journal of transportation science and technology, 7(2), 151-167.
80. Wu, X., Feng, Q., Bai, C., Lai, C. S., Jia, Y., Lai, L. L. (2021). A novel fast-charging stations locational planning model for electric bus transit system. Energy, 224, 120106.
81. Xiao, Y., Zhang, Y., Kaku, I., Kang, R., Pan, X. (2021). Electric vehicle routing problem: A systematic review and a new comprehensive model with nonlinear energy recharging and consumption. Renewable and Sustainable Energy Reviews, 151, 111567.
82. Yi, Z., Smart, J. (2021). A framework for integrated dispatching and charging management of an autonomous electric vehicle ride-hailing fleet. Transportation Research Part D: Transport and Environment, 95, 102822.
83. Yu G., Liu A., Zhang J., Sun H. (2021). Optimal operations planning of electric autonomous vehicles via asynchronous learning in ride-hailing systems. Omega, 102448.
84. Zalesak, M., Samaranayake, S. (2021). Real Time Operation of High-Capacity Electric Vehicle Ridesharing Fleets. https://arxiv.org/abs/2101.01274.
85. Zhang H, Sheppard CJ, Lipman TE, Moura SJ, (2020). Joint fleet sizing and charging system planning for autonomous electric vehicles. IEEE Trans. Intell. Transp. Sys. 21(11), 4725 – 4738.
86. Zhang, Y., Lu, M., Shen, S. (2021). On the values of vehicle-to-grid electricity selling in electric vehicle sharing. Manufacturing & Service Operations Management, 23(2), 488-507.
87. Zhu, M., Liu, X. Y., Wang, X. (2018). Joint transportation and charging scheduling in public vehicle systems—a game theoretic approach. IEEE Transactions on Intelligent Transportation Systems, 19(8), 2407-2419.
88. Ziad, C., Rajamani, H.S., Manikas, I., (2019). Game-theoretic Approach to Fleet Management for Vehicle to Grid Services. 2019 IEEE 19th Int. Symp. Signal Process. Inf. Technol. ISSPIT 2019.